\documentclass[a4paper,oneside,final,onecolumn,12pt]{article}
\usepackage{amssymb}

\newcounter{subequation}

\input{tcilatex}

\begin{document}

\title{Mean Field critical behaviour for a Fully Frustrated Blume-Emery-Griffiths
Model}
\author{Francesco di Liberto\thanks{%
Corresponding author.Tel+39-81-7253424; e-mail: diliberto@na.infn.it} \ and
Fulvio Peruggi \\
\\
{\small \textit{INFM and INFN Sezione di Napoli}}\\
{\small \textit{Dipartimento di Scienze Fisiche, Universit\`{a} di Napoli
``Federico II''}} \\
{\small \textit{Complesso Universitario di Monte S. Angelo}}\\
{\small \textit{\ Via Cintia Lotto G. I-80125 Napoli, Italy}}}
\maketitle

\begin{abstract}
We present a mean field analysis of a fully frustrated Ising spin model on
an Ising lattice gas. This is equivalent to a degenerate
Blume-Emery-Griffiths model with frustration, which we analyze for different
values of the quadrupolar interaction.This model might be useful in the
study of structural glasses and related systems with disorder.
\end{abstract}

\date{PACS: 05.50 }

\section{Introduction}

In the last two decades\ the physics of complex systems, ranging from\
dilute magnets to structural glasses has been captured by models which
couple Ising variables with lattice gas or Potts variables\ \cite{1}-\cite{6}%
, i.e. models with this type of Hamiltonian:

\begin{equation}
-\beta \mathcal{H}=\sum_{\langle ij\rangle }J\varepsilon
_{ij}S_{i}S_{j}n_{i}n_{j}+\sum_{\langle ij\rangle }Kn_{i}n_{j}+\mu
\sum_{i}n_{i},  \label{1}
\end{equation}
where $\varepsilon _{ij}=\pm 1$ are quenched variables associated to pairs
of nearest neighbour sites, $J>0$ is the interaction between the Ising spin
variables$\;(S=\pm 1),$ $K$ is the interaction between the particles,\ $%
n_{i}=0,1$\ are the lattice gas variables, $\mu $ is the chemical potential.
The spins can interact\ each other ferromagnetically ($\varepsilon _{ij}=1$)
or antiferromagnetically ($\varepsilon _{ij}=-1$).

For $\varepsilon _{ij}=1$ everywhere, this model concides with the original
Blume-Emery-Griffiths model (BEG) \cite{7}-\cite{17} with an extra
degeneracy\ 2 at each empty site. $J\varepsilon _{ij}$ is the bilinear
interaction, $K$ the quadrupolar interaction, and $\mu $ the crystal field.
In the last few years the disordered BEG model has been studied for random
values of the $\varepsilon _{ij}=\pm 1$ \cite{18,19}.\ Recently the
Degenerate BEG (DBEG) \cite{20,21} has been found suitable to describe the
martensitic trasformation.

It may be useful to write the Hamiltonian of Eq.\ (\ref{1}) in the following
way:

\begin{equation}
-\beta \mathcal{H}=\sum_{\langle ij\rangle }[J(\varepsilon
_{ij}S_{i}S_{j}-1)n_{i}n_{j}+\eta Jn_{i}n_{j}]+\mu \sum_{i}n_{i},
\label{2bis}
\end{equation}
where $\eta =K/J+1.$ For $\eta =0$\ and $J=\infty ,$ this model has been
extensively studied in the last few years to study glassy systems and
granular materials in\ the disordered case (i.e. when the $\varepsilon _{ij}$
variables are randomly distributed on the lattice ) \cite{6}, \cite{22}-\cite
{29}. This model can be considered as a model of particles with an internal
degree of freedom ($S=\pm 1)$ that interact with an effective coupling $%
J(\varepsilon _{ij}S_{i}S_{j}-1)$ which is zero for spin configurations that
satisfy the interactions (i.e. $\varepsilon _{ij}S_{i}S_{j}=1)$ and gives an
infinite repulsion, for those that do not satisfy the interaction (i.e. $%
\varepsilon _{ij}S_{i}S_{j}\neq 1)$. So these last configurations are
forbidden for $J=\infty $.

Here we analyze the model for $J$ finite and $\eta \geqslant 0.$ For $\eta
\neq 0$\ there is an extra interaction between a pair of n.n.
particles,while finite values of $J$\ correspond to softening the hard core
potential between the spin variables.

In particular we present a mean field analysis of the Hamiltonian of Eq.\ (%
\ref{1}) in the fully frustrated (FF) case on the square lattice. In this
case the $\varepsilon _{ij}$ variables are choosen in such a way that every
plaquette (i.e.\ elementary cell of the lattice) is frustrated. In other
terms every plaquette has an odd number of $\varepsilon _{ij}=-1$, so that
the four spins of the plaquette cannot completely satisfy the interactions.
In Fig.\ 1 we show the Villain \cite{30} scheme for the 2D FF model,
highlighting the differences between the A and B sublattices. For this FF
lattice we have recently \cite{31} made a mean field analysis of the
Frustrated Percolation problem \cite{32}-\cite{43}.

In Sec. 3\ and Sec. 4 we write down the equations for site magnetizations $%
(m_{A}$ and $m_{B})$ and site densities ($D_{A}$ and $D_{B})$ and these
enable us to find the critical lines for the order-disorder transitions in
our model for the FF case.

For $K/J$ $>-1$ (i.e. $\eta >0)$ there is a tricritical point which
separates the critical line in two branches, respectively characterized by
first-order and second-order transitions. On the other hand for $K/J=-1\;($%
i.e. $\eta =0)$ the transitions are second-order for any $\mu .$

Finally we compare the FF behaviour with that of the original Ferromagnetic
BEG with and without degeneracy.

\section{Mean field analysis}

We will study the model defined by the Hamiltonian (\ref{1}) by evaluating
its free energy in a mean field approximation. For convenience we will set $%
\kappa =K/J$

At each site $i$ of the lattice we have to consider the variables $S_{i} =
\pm 1$ and $n_{i} = 0,1$. For notation purposes it is useful to introduce a
new 4-state variable $\nu_{i}$ such that $\left\{ \nu_{i} \right\} = \left\{
n_{i} \right\} \otimes \left\{ S_{i} \right\} = \left\{ 1\uparrow,
1\downarrow, 0\uparrow, 0\downarrow \right\} \equiv \left\{ 1, 2, 3, 4
\right\}$. We can express the old variables in terms of this new variable by
means of the relations: $n_{i} S_{i} = \delta_{\nu_{i},1} -
\delta_{\nu_{i},2}$ and $n_{i} = \delta_{\nu_{i},1} + \delta_{\nu_{i},2}$.

Moreover, using the index $r$ to denote one of the four states of $\nu_{i}$,
we can define $p_{r}^{i} = \left\langle \delta_{\nu_{i},r} \right\rangle$,
i.e.\ the probability that the site $i$ will be found in the state $%
\nu_{i}=r $. Here the angular brackets represent, as usual, the average done
with the Hamiltonian of Eq.\ (\ref{1}).

To obtain the free energy we evaluate first the internal energy of the
system, which is the expectation value of our Hamiltonian: 
\begin{eqnarray}
-\beta \mathcal{U}\; &\equiv &\left\langle -\beta \mathcal{H}\right\rangle
=-\beta \sum_{\langle ij\rangle }\left\langle \mathcal{H}_{ij}\right\rangle
-\beta \sum_{i}\left\langle \mathcal{H}_{i}\right\rangle =  \nonumber \\
&=&J\sum_{\left\langle ij\right\rangle }\varepsilon _{ij}\left\langle
(\delta _{\nu _{i},1}-\delta _{\nu _{i},2})(\delta _{\nu _{j},1}-\delta
_{\nu _{j},2})\right\rangle + \\
&&+\kappa \sum_{\left\langle ij\right\rangle }\left\langle (\delta _{\nu
_{i},1}+\delta _{\nu _{i},2})(\delta _{\nu _{j},1}+\delta _{\nu
_{j},2})\right\rangle +\mu \sum_{i}\left\langle \delta _{\nu _{i},1}+\delta
_{\nu _{i},2}\right\rangle .  \nonumber
\end{eqnarray}
In the MF context we neglect the fluctuations and can simply put 
\begin{equation}
\left\langle \delta _{\nu _{i},r}\delta _{\nu _{i},s}\right\rangle
=\left\langle \delta _{\nu _{i},r}\right\rangle \left\langle \delta _{\nu
_{i},s}\right\rangle ,  \label{3}
\end{equation}
so relation (\ref{3}) implies \renewcommand{\theequation}{\arabic{equation}%
\alph{subequation}} 
\begin{eqnarray}
\setcounter{subequation}{0}\setcounter{subequation}{1}\left\langle -\beta 
\mathcal{H}_{ij}\right\rangle &=&J\left[ \varepsilon _{ij}\left(
p_{1}^{i}-p_{2}^{i}\right) \left( p_{1}^{j}-p_{2}^{j}\right) +\kappa \left(
p_{1}^{i}+p_{2}^{i}\right) \left( p_{1}^{j}+p_{2}^{j}\right) \right] ,
\label{3'a} \\
\addtocounter{equation}{-1}\setcounter{subequation}{2}\left\langle -\beta 
\mathcal{H}_{i}\right\rangle &=&\mu \left( p_{1}^{i}+p_{2}^{i}\right) .
\label{3'b}
\end{eqnarray}
\renewcommand{\theequation}{\arabic{equation}}

The order parameters we will use in the following are the site magnetization 
$m_{i}$ and the lattice gas particle density $D_{i}$ expressed by %
\renewcommand{\theequation}{\arabic{equation}\alph{subequation}} 
\begin{eqnarray}
\setcounter{subequation}{0}\setcounter{subequation}{1}m_{i} &=&\left\langle
S_{i}n_{i}\right\rangle =\left\langle \delta _{\nu _{i},1}-\delta _{\nu
_{i},2}\right\rangle =p_{1}^{i}-p_{2}^{i},  \label{4a} \\
\addtocounter{equation}{-1}\setcounter{subequation}{2}D_{i} &=&\left\langle
n_{i}\right\rangle =\left\langle \delta _{\nu _{i},1}+\delta _{\nu
_{i},2}\right\rangle =p_{1}^{i}+p_{2}^{i},  \label{4b}
\end{eqnarray}
\renewcommand{\theequation}{\arabic{equation}} from which we have: %
\renewcommand{\theequation}{\arabic{equation}\alph{subequation}} 
\begin{equation}
\setcounter{subequation}{1}p_{1}^{i}=\frac{1}{2}(D_{i}+m_{i})\qquad
p_{2}^{i}=\frac{1}{2}(D_{i}-m_{i}).  \label{5a}
\end{equation}
\renewcommand{\theequation}{\arabic{equation}}

These relations and the equivalence condition $p_{3}^{i}=p_{4}^{i}$,
together with the normalization $\sum_{r=1}^{4}p_{r}^{i}=1$, imply:%
\renewcommand{\theequation}{\arabic{equation}\alph{subequation}} 
\begin{equation}
\addtocounter{equation}{-1}\setcounter{subequation}{2}p_{3}^{i}=p_{4}^{i}=%
\frac{1}{2}(1-D_{i}).  \label{5b}
\end{equation}
\renewcommand{\theequation}{\arabic{equation}}

Moreover we invoke the typical translation invariance requirement of the MF
approximation, taken separately on the two sublattices. Then we look for a
solution in which all the sites of sublattice A (B) have the same
probabilities, i.e.\ $p_{r}^{i} = p_{r}^{A} \ \forall\ i \in A$ and $%
p_{r}^{i} = p_{r}^{B} \ \forall\ i \in B$. This solution is one of the many
occurring in the degenerate ground state.

Using the translation invariance we can write 
\[
-\beta \mathcal{H}_{AB}=J\left[ m_{A}m_{B}+\kappa D_{A}D_{B}\right] 
\]
for the expectation value $\left\langle -\beta \mathcal{H}_{ij}\right\rangle 
$ of the partial Hamiltonian relative to any AB ferromagnetic bond, i.e.\
any ferromagnetic bond $\langle ij\rangle $ such that $i\in A$ and $j\in B$.
A similar relation holds for all the partial Hamiltonians relative to any AA
ferromagnetic bond. On the other hand, the expectation value of the partial
Hamiltonian relative to any BB antiferromagnetic bond ($\varepsilon
_{ij}=-1) $ is given by 
\[
-\beta \mathcal{H}_{BB}=J\left[ -m_{B}^{2}+\kappa D_{B}^{2}\right] . 
\]
Therefore, for $N$ sites, since the number of A sites and the number of B
sites are both $N/2$, the internal energy is 
\begin{eqnarray}
\frac{-\beta \mathcal{U}}{N} &=&\frac{1}{N}\sum_{i=1}^{N}\left[ \frac{1}{2}%
\sum_{j:\exists \left\langle ij\right\rangle }\left\langle -\beta \mathcal{H}%
_{ij}\right\rangle +\mu D_{i}\right]  \nonumber  \label{6} \\
&=&\frac{1}{N}\sum_{i\in A}\frac{1}{2}\left\{ \frac{z}{2}\left\langle -\beta 
\mathcal{H}_{AA}\right\rangle +\frac{z}{2}\left\langle -\beta \mathcal{H}%
_{AB}\right\rangle +\mu D_{A}\right\}  \nonumber \\
&&+\frac{1}{N}\sum_{i\in B}\frac{1}{2}\left\{ \frac{z}{2}\left\langle -\beta 
\mathcal{H}_{BA}\right\rangle +\frac{z}{2}\left\langle -\beta \mathcal{H}%
_{BB}\right\rangle +\mu D_{B}\right\}  \nonumber \\
\ &&  \nonumber \\
&=&\frac{Jz}{8}\left[ m_{A}^{2}+2m_{A}m_{B}-m_{B}^{2}+\kappa \left(
D_{A}+D_{B}\right) ^{2}\right] +\frac{1}{2}\mu \left( D_{A}+D_{B}\right) . 
\nonumber \\
\ &&
\end{eqnarray}

For the evaluation of the MF entropic term we use the factorization property
of the probability distribution $\mathcal{P}(\nu_{1}, \ldots, \nu_{N})$ and
therefore get $\mathcal{S} \equiv -k \sum_{\left\{ \nu \right\} } \mathcal{P}
\ln \mathcal{P} = -k \sum_{i=1}^{N} \sum_{r=1}^{4} p_{r}^{i} \ln p_{r}^{i}$.

Using the translation invariance, this can be written in the form 
\begin{equation}  \label{7}
\frac{\mathcal{S}}{kN} = - \frac{1}{2} \sum_{r=1}^{4} (p_{r}^{A} \ln
p_{r}^{A} + p_{r}^{B} \ln p_{r}^{B}).
\end{equation}

Using Eqs.\ (\ref{6}) and (\ref{7}) we can finally write the MF free energy
per site of the lattice: 
\begin{equation}
\beta f\equiv \frac{\beta \mathcal{F}}{N}\equiv \frac{\beta \mathcal{U}}{N}-%
\frac{\mathcal{S}}{kN},  \label{8}
\end{equation}
where the probabilities $p_{r}^{A}$ and $p_{r}^{B}$ have to be expressed in
terms of the local order parameters $m_{A},m_{B},D_{A}$ and $D_{B}$ through
Eq (7).

\section{ Equations for the site Magnetizations and Densities}

The knowledge of the free energy allows us to write down easily the MF
equations that must be satisfied by the order parameters $m_{A}$, $m_{B}$, $%
D_{A}$ and $D_{B}$.

From the stationary relations $\partial f/\partial m_{A}=0$ and $\partial
f/\partial m_{B}=0$ it follows that 
\begin{equation}
m_{A}=D_{A}\tanh \left( \frac{\lambda }{2}(m_{A}+m_{B})\right) ,\qquad
m_{B}=D_{B}\tanh \left( \frac{\lambda }{2}(m_{A}-m_{B})\right) .  \label{9}
\end{equation}
Here $\lambda =4J=4J_{o}/kT=T_{c}/T$ where $T_{c}$ $\equiv 4J_{0}/k$ is the
mean field critical temperature of the isotropic Ising model recovered by
the isotropic version of the Hamiltonian (\ref{1}) in the $\mu \rightarrow
\infty $ limit.

Morover from the stationary relations $\partial f/\partial D_{A}=0$ and $%
\partial f/\partial D_{B}=0$ we deduce that 
\begin{eqnarray}  \label{10}
e^{\kappa \lambda (D_{A}+D_{B}) + 2\mu} & = & \frac{ D_{A}^{2}-m_{A}^{2} }{
(1-D_{A})^{2} },  \nonumber \\
& & \\
e^{\kappa \lambda (D_{A}+D_{B}) + 2\mu} & = & \frac{ D_{B}^{2}-m_{B}^{2} }{
(1-D_{B})^{2} }.  \nonumber
\end{eqnarray}
\renewcommand{\theequation}{\arabic{equation}} These relations give in
implicit form $D_{A}$ and $D_{B}$ for every $m_{A}$ and $m_{B}.$

Now, replacing Eqs.\ (\ref{9}) into Eqs.\ (\ref{10}) we get stationarity in
the four order parameters $m_{A}$, $m_{B}$, $D_{A}$ and $D_{B}$. After
straightforward calculations we find:\renewcommand{\theequation}{%
\arabic{equation}\alph{subequation}} 
\begin{eqnarray}
\setcounter{subequation}{0}\setcounter{subequation}{1}D_{A} &=&\frac{\cosh %
\left[ (\lambda /2)(m_{A}+m_{B})\right] }{e^{(-\kappa \lambda
/2)(D_{A}+D_{B})-\mu }+\cosh \left[ (\lambda /2)(m_{A}+m_{B})\right] },
\label{11a} \\
\addtocounter{equation}{-1}\setcounter{subequation}{2}m_{A} &=&\frac{\sinh %
\left[ (\lambda /2)(m_{A}+m_{B})\right] }{e^{(-\kappa \lambda
/2)(D_{A}+D_{B})-\mu }+\cosh \left[ (\lambda /2)(m_{A}+m_{B})\right] },
\label{11b} \\
\setcounter{subequation}{0}\setcounter{subequation}{1}D_{B} &=&\frac{\cosh %
\left[ (\lambda /2)(m_{A}-m_{B})\right] }{e^{(-\kappa \lambda
/2)(D_{A}+D_{B})-\mu }+\cosh \left[ (\lambda /2)(m_{A}-m_{B})\right] },
\label{11c} \\
\addtocounter{equation}{-1}\setcounter{subequation}{2}m_{B} &=&\frac{\sinh %
\left[ (\lambda /2)(m_{A}-m_{B})\right] }{e^{(-\kappa \lambda
/2)(D_{A}+D_{B})-\mu }+\cosh \left[ (\lambda /2)(m_{A}-m_{B})\right] }.
\label{11d}
\end{eqnarray}
\renewcommand{\theequation}{\arabic{equation}}

These equations can be studied numerically for different values of $\kappa $
in order to find the fixed points for every $\lambda $ and $\mu $. This
analysis, together with the values of the free energy (\ref{8}) for each
fixed point, has enabled us to find for every $\mu $ the critical value $%
\lambda _{c}$ where the order parameters $m_{A}$ , $m_{B}$ , $D_{A}$ and $%
D_{B}$ undergo a first-order or second-order transition.

\section{Critical lines and Results}

We have done our analysis for a number of values of the $\kappa $ parameter,
but report here, for convenience, only the most interesting cases in the
range$\ \kappa \geqslant -1$ (i.e. $\eta \geqslant 0)$. Note that the
antiquadrupolar phase that generally appears in the BEG model for $\kappa <0$
does not appear here because our sublattice partition is intrinsically
different from the usual BEG sublattice partition.\textit{\ }The critical
behaviours are reported in Fig. 2--6 respectively for $\eta =1.16,$ $1,$ $%
.84,$ $.5,$ $0.$ To appreciate the differences between the FF model and the
Ferromagnetic model (i.e.\ $\varepsilon _{ij}=1$ for all bonds), each figure
contains the (a)-section in which we report the behaviour of the Degenerate
FF BEG model and the (b)-section relative to behaviour of Degenerate
Ferromagnetic BEG model. In the (a)-section for each $T/T_{c}$ we give the
field $-\mu /\lambda $ were the transition from the high-field disordered
phase ($m_{A}=m_{B}=0$ and $D_{A}=D_{B}$) to the low-field ordered phase ($%
m_{A}>m_{B}\neq 0$ and $D_{A}>D_{B}$) takes place. Bold (dotted) lines
represents second-order (first-order) transitions. Dashed lines represent
the spinodals, i.e.\ the boundaries of areas of metastability that surround
any first-order transition line. Below the first-order transition line. the
metastable phase is the disordered phase, above this line the metastable
phase is the ordered phase \cite{44}. In the (b)-section for each $T/T_{c}$
we give the field $-\mu /\lambda $ were the transition from the high-field
disordered phase ($m=0$ and $D\leq 1/2$) to the low-field ordered phase ($%
m>0 $ and $D\geq 1/2)$ takes place. As for the (a)-section, bold (dotted)
lines represents second-order (first-order) transitions; dashed lines
represent the spinodals. Fig. 2-6 is relative to decreasing values\ of the
extra-interaction $\eta =\kappa +1.$ The overall feature is that decreasing $%
\eta $ we obtain a smaller ordered region. This\ is expected if we look at
the Hamiltonian (\ref{2bis}) since $\eta $ is the extra interaction among
the particles. In the insert of Fig. 3b and Fig. 4b we report also the
behaviour of the original BEG.

For the Ferromagnetic Degenerate BEG\ we find that the degeneracy reduces
the area of the ordered region and increases the area of the region of
first-order transitions, in agreement with recent results \cite{20,21}.

On the other hand it is known that the frustration has the conflicting
effect of reducing this region both for the original BEG with random bonds 
\cite{12} and for the DBEG with random field \cite{21}. Here we find that
the frustration reduces the ordered region and moves the tricritical point
toward low temperatures, i.e. the frustration in the Fully-Frustrated model
(in spite of the small degeneracy present) reduces the first order region.

These results may be useful to study the effects of the softening of the
hard core potential and the effect of the attraction between particles for
systems described by Hamiltonian (\ref{2bis}) such as for example glasses
and granular material.\bigskip

\bigskip \bigskip \noindent \textbf{\Large Acknowledgements} \bigskip

We gratefully aknowledge A.\ Coniglio and A.\ De Candia for useful
discussions.

\pagebreak \noindent\textbf{\Large Figure captions}

\begin{description}
\item[\textbf{Figure 1}]  $2d$ FF model on the square lattice. Straight
(wavy) lines represent ferromagnetic (antiferromagnetic) interactions. $z=4$
ferromagnetic interactions start from each site of the sublattice $A$ (open
circles); $z/2$ ferromagnetic interactions and $z/2$ antiferromagnetic
interactions start from each site of the sublattice $B$ (closed circles).

\item[\textbf{Figure 2}]  (a) Critical lines for the FF lattice for $\kappa
=+.16\;($i.e. $\eta =1.16)$. Bold (dotted) lines represents second-order
(first-order) transitions. Dashed lines represent the spinodals. (b)
Corresponding critical lines for the ferromagnetic model (i.e.\ $\varepsilon
_{ij}=1$ for all bonds).

\item[\textbf{Figure 3}]  (a) Critical lines for the FF lattice for $\kappa
=0$ \ ($\eta =1)$. The tricritical point is located at $T/T_{c}\approx 0,233$
and $-\mu /\lambda =(1/\lambda )\ln \left( -1+\lambda /\sqrt{2}\right)
\approx .166$. (b) Corresponding critical lines for the ferromagnetic model.
The insert reports the critical lines for the original BEG \cite{7}.

\item[\textbf{Figure 4}]  (a) Critical lines for the FF lattice for $\kappa
=-.16\;(\eta =.84)$., (b) Corresponding critical lines for the ferromagnetic
model Ferromagnetic. In the insert we report the corresponding critical
lines for the original BEG \cite{7}.

\item[\textbf{Figure 5}]  (a) Critical lines for the FF lattice for $\kappa
=-.5\;(\eta =+.5)$, (b) Corresponding critical lines for the ferromagnetic
model. Observe that both in the ferromagnetic and fully-frustrated case the
first-order transition line continues in the ordered phase, below the
tricrical point,similarly to the corresponding behaviour of the original BEG 
\cite{14,15}.

\item[\textbf{Figure 6}]  (a) Critical lines for the FF lattice for $\kappa
=-1\;(\eta =0)$, (b) Corresponding critical lines for the ferromagnetic
model. Observe that the first-order transition line now disappears,
differently from what happens in the spin glass case \cite{18,19}.
\end{description}

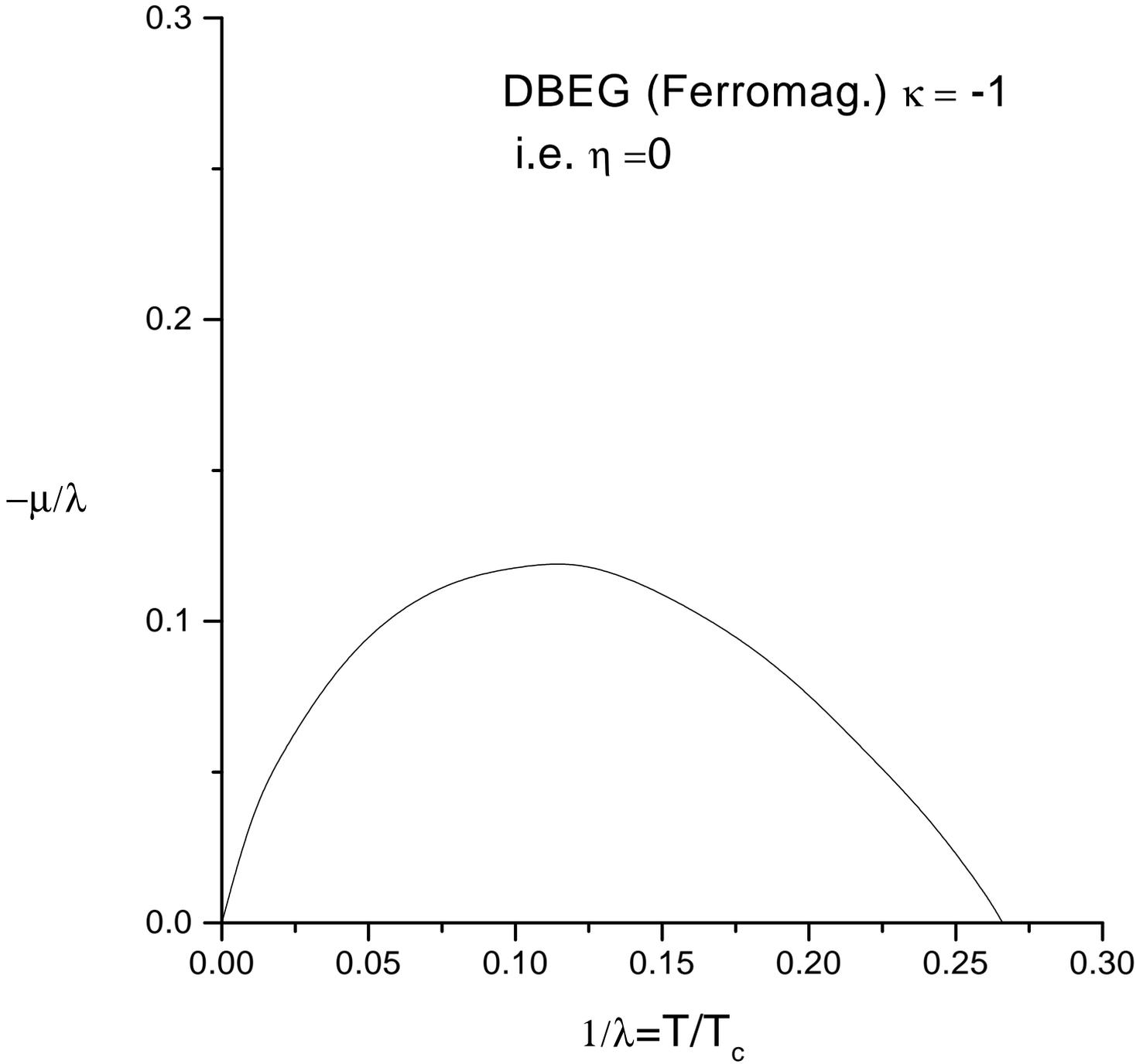
\end{document}